# Effects of Treatment on the Treated: Identification and Generalization

**Ilya Shpitser**
Department of Epidemiology
Harvard School of Public Health
ishpitse@hsph.harvard.edu

**Judea Pearl**
Department of Computer Science
University of California, Los Angeles
judea@cs.ucla.edu

## Abstract

Many applications of causal analysis call for assessing, retrospectively, the effect of withholding an action that has in fact been implemented. This counterfactual quantity, sometimes called "effect of treatment on the treated," (ETT) have been used to to evaluate educational programs, critic public policies, and justify individual decision making. In this paper we explore the conditions under which ETT can be estimated from (i.e., identified in) experimental and/or observational studies. We show that, when the action invokes a singleton variable, the conditions for ETT identification have simple characterizations in terms of causal diagrams. We further give a graphical characterization of the conditions under which the effects of multiple treatments on the treated can be identified, as well as ways in which the ETT estimand can be constructed from both interventional and observational distributions.

## 1 Introduction

Consider the following two problems, involving personal decision and program evaluation, respectively.

The first concerns a painful treatment that consists of administering a dose $x'$ of a serum to the bloodstream of patients who are judged to need such treatment. A patient who agreed to take the recommended treatment is wondering whether it was worth the discomfort, namely, what his chances of recovery would have been had he taken a lower dose of serum, say $x$. This counterfactual quantity seems to defy empirical measurement because we can never rerun history and administer a different level of treatment for those who already received it at level $x'$; the latter might possess distinct needs and dispositions that make them react differently to a reduced dosage $x$ than a randomly selected patient would.

The second problem involves a policy maker considering the termination of an ongoing job-training program, seeking to estimate the anticipated reduction in future earning of those enrolled in the program. This calls for comparing the future earning of the program's graduates to their hypothetical earning had they not been trained. Again, because those who enroll in the program have special needs and qualities, comparison to the population at large will not be adequate.

In both examples, we ask for the value of an outcome variable $Y$ after setting the value of another variable $X$ to $x$, but also knowing that $X$ attains value $x'$ naturally. Letting $Y_x$ be a random variable representing the behavior of $Y$ after $X$ is set to value $x$, each of the two queries can be represented by the expression $P(Y_x = y|x')$, often called the *effect of treatment on the treated* (ETT) [1]. Modeling and identification issues which arise in these queries have received some attention in the literature [6]. In subsequent sections, we introduce the machinery of causal inference necessary to define such expressions precisely using *causal diagrams* as carriers of background knowledge, and show that, if $X$ is a singleton and **Y** is a set, ETT can be computed from the observational distribution if the causal diagram satisfies a simple graphical criterion. Finally, we show that if **X** is a set, the identification of ETT hinges on a more complex graphical criterion which, nevertheless, yields a simple way of constructing the ETT estimand.

## 2 Preliminaries

In this paper, we formalize counterfactual expressions using the Structural Causal Model (SCM) defined in [4], Chapter 7. Such models have a marked advantage over the potential-outcome approach of [2] and [7] in that they permit background knowledge to be expressed in the ordinary scientific language of cause-effect relationships instead of the artificial language of counterfactual independencies required by the latter. The SCM consists of a set of observed variables $\mathbf{V}$ and unobserved or latent variables $\mathbf{U}$. Without loss of generality, it is usually assumed that the unobserved variables are exogenous and random, and the observed variables are endogenous, whose values are obtained from the values of other variables by means of unknown functions. Unobserved variables are drawn from a joint distribution $P$, and this distribution, together with the functional mapping onto the endogenous variables, defines an observed distribution $P(\mathbf{v})$.

Each causal model induces a directed graph, called a causal diagram. In such a graph, each variable in model is represented by a vertex, and a vertex corresponding to a variable $V_i$ has arrows incoming from every variable whose value is used to determine the value of $V_i$ by its determining function. In such a graph exogenous variables have no incoming arrows, e.g. have no parents in the graph. In this paper we restrict our attention to acyclic causal diagrams. We will use the standard graph-theoretic family abbreviations for graph relations, e.g. $An(\mathbf{X})_G$, $De(\mathbf{X})_G$, $Pa(\mathbf{X})_G$, $Ch(\mathbf{X})_G$ stand for the set of ancestors, descendants, parents and children of $\mathbf{X}$ in $G$, not including $\mathbf{X}$.

The value of causal diagrams is their ability to display conditional independence among variables in terms of path-separation criterion known as d-separation [3]. Two sets of nodes $\mathbf{X}, \mathbf{Y}$ are said to be d-separated by a third set $\mathbf{Z}$ if every edge path from nodes in one set to nodes in another are "blocked" where blockage occurs when one of the following triples occurs on the path: $X \to Z \to Y$, $X \leftarrow Z \to Y$, and $X \to W \leftarrow Y$, where $Z \in \mathbf{Z}$, and neither $W$ nor any descendant of $W$ is in $\mathbf{Z}$. Paths (or node sets) that are not d-separated are called d-connected. A d-connected path starting with an outgoing arrow is called a frontdoor path, while a d-connected path starting with an incoming arrow is called a backdoor path.

If many of the variables in a causal diagram are latent, it can be inconvenient to consider long path stretches containing nothing but latent variables. One graphical representation that avoids this is the *latent projection* [14]. A latent projection of a causal diagram is a mixed graph containing directed and bidirected arcs, where there is a vertex for every observable node, and two observable nodes $X, Y$ are connected by a directed arrow if there is a d-connected path from $X$ to $Y$ in the original causal diagram containing only arrows pointing away from $X$ and towards $Y$, and each node on this path other than $X$ and $Y$ is latent. Similarly, $X$ and $Y$ are connected by a bidirected arrow if there is a d-connected path from $X$ to $Y$ in the original causal diagram which starts with an arrow pointing to $X$ and ends in an arrow pointing to $Y$. D-separation generalizes in a straightforward way to latent projections [5], and latent projections preserve all d-separation statements of the original causal diagrams. In the remainder of the paper we will use latent projections, and refer to them as graphs or causal diagrams.

Since endogenous variables in a causal model are causally determined via functions, causal diagrams or alternative representations such as latent projections encode more than just conditional independence statements, they also encode direct causal relationships between variables. This additional layer of meaning allows causal diagrams to represent not only probabilistic operations such as marginalization or conditioning, but causal operations such as interventions.

An *intervention*, denoted by $do(\mathbf{x})$ in [4], is an operation where values of a set of variables $\mathbf{X}$ are set to $\mathbf{x}$ without regard of how values of $\mathbf{X}$ are ordinarily determined in the model via functions. The responses of the remaining observable variables other than $\mathbf{X}$ to the intervention are represented by an interventional distribution denoted as $P(\mathbf{v} \setminus \mathbf{x} | do(\mathbf{x}))$ or $P_\mathbf{x}(\mathbf{v} \setminus \mathbf{x})$. The response of a single observable variable $Y$ to $do(\mathbf{x})$ is sometimes denoted by a counterfactual variable $Y_\mathbf{x}$.

Once we fix the value of every exogenous variable in the model, the remaining variables become deterministically fixed. This allows us to use the distribution $P$ over exogenous variables to define joint distributions over counterfactual variables, even if the interventions which determine these variables disagree with each other. In other words,

$$P(Y_{\mathbf{x}^1}^1 = y^1, ..., Y_{\mathbf{x}^k}^k = y^k) = \sum_{\{\mathbf{e} | Y_{\mathbf{x}^1}^1(\mathbf{e}) = y^1 \wedge ... \wedge Y_{\mathbf{x}^k}^k(\mathbf{e}) = y^k\}} P(\mathbf{e})$$

where $\mathbf{E}$ is the set of exogenous variables in the model.

Note that the queries mentioned in the introduction can all be represented by the expression $P(Y_x = y | X = x')$, which is a conditional distribution obtainable from a counterfactual joint distribution of the type defined above. In practice, we cannot use this definition to evaluate these distributions directly since some exogenous variables may not be observed, in which case $P(\mathbf{e})$ cannot be estimated, and functions which determine various variables are not generally known either, which means evaluating joint response of the same variable to different interventions becomes difficult.

Furthermore, it's unclear how to design an experimental study to estimate counterfactual quantities. This is because counterfactuals of the kind we are interested in involve conflicts in value assignments which do not correspond in an obvious way to any exper-imental regime we can impose. An alternative way of obtaining estimates of counterfactual quantities is to express them in terms of the observable distribution $P(\mathbf{v})$, given causal and probabilistic constraints encoded by a causal diagram. A counterfactual $\gamma$ expressible in this way is called identifiable from $P(\mathbf{v})$ in causal diagram $G$, (written $P(\mathbf{v}), G \vdash_{id} \gamma$), and determining which queries $\gamma$ are expressible in this way is known as the identification problem [4].

## 3 Effects of Single Treatment on the Treated

In this section, we consider identification of queries of the form $P(Y_x^1 = y^1, ..., Y_x^k = y^k | X = x')$, which we will abbreviate as $P(\mathbf{Y}_x = \mathbf{y} | x')$, where $\mathbf{Y} = \{Y^1, ..., Y^k\}$. We will assume $X$ has a causal influence on some outcomes, in other words $X \in An(\mathbf{Y})_G$. The identification of causal effect queries, $P(\mathbf{y}|do(\mathbf{x}))$, require two graphical models, one governing the pre-intervention distribution, and one governing the post-intervention distribution. The latter is represented by the original causal diagram with all arrows pointing towards $\mathbf{X}$ removed, since interventions sever the influence of the parents on $\mathbf{X}$. We will denote such a graph obtained from $G$ as $G_{\overline{x}}$.

Counterfactual queries involve a mixture of events representing multiple hypothetical worlds, where each world corresponds to an interventional regime of a single counterfactual antecedent, and is represented by a copy of the original causal diagram, with the appropriate arrows removed. Furthermore, each of these worlds are assumed to share history up until the intervention. This sharing is modeled by each copy sharing the exogenous variables. If the values of some observable variables are known or observed, some otherwise distinct vertices in different copies may end up being the same due to the axiom of composition [4]. The resulting graph containing these copies is known as a counterfactual graph [10].

Causal effect expressions, e.g., $P(y|do(x)) = P(Y_x = y)$ and in fact all expressions in do-calculus can be evaluated from experimental studies because they belong to a class of counterfactual sentences in which all quantities refer to the same hypothetical world, specified by the counterfactual antecedent. For example, $P(y|w, do(x), do(z))$ translates to the counterfactual $P(Y_{xz} = y | W_{xz} = w)$, in which all subscripts are the same. The ETT expression $P(Y_x = y | X = x')$, on the other hand, involves quantities in two different worlds; a post-treatment quantity $Y_x$ and a pre-treatment quantity $X$ – the latter changed by treatment.

For this expression, the counterfactual graph is fairly straightforward. One possible world, corresponding to variables $Y_x^i$, will be represented by the causal diagram with arrows incoming to $X$ cut. Another possible world, corresponding to the variable assignment $X = x'$, will contain the original causal diagram. Since our causal diagrams are latent projections, we can assume without loss of generality that all exogenous variables shared by these two worlds are latent. Furthermore, because we aren't interested in any descendant of $X$ in the second world, no such nodes need to be added to the graph representing the second world. Finally, since intervening on $X$ cannot influence non-descendants of $X$, the non-descendants of $X$ in both of the possible worlds are the same random variables, and so can be merged. As a result, we can prove the following theorem.

**Theorem 1** $P(\mathbf{Y}_x = \mathbf{y}|x')$ *is identifiable in $G$ if and only if $P(\mathbf{y}|w, do(x))$ is identifiable in $G'$, where $G'$ is obtained from $G$ by adding a new node $W$ with the same set of parents (both observed and unobserved) as $X$, and no children. Moreover, the estimand for $P(\mathbf{Y}_x = \mathbf{y}|x')$ is equal to that of $P(\mathbf{y}|w, do(x))$ with all occurrences of $w$ replaced by $x'$.*

*Proof:* Using above reasoning, we established that the counterfactual graph for identifying $P(\mathbf{Y}_x = \mathbf{y}|x')$ contains a copy of $G$ with the arrows into $X$ removed, and another copy of $G$, and that the non-descendants of $X$ in both copies are shared, while descendants of $X$ in the second copy can be marginalized away. The resulting graph is precisely $G'$.

First we show the if and only if part of the theorem. $W$ either has a backdoor path to $\mathbf{Y}$ or not. If it does not, then $W$ is d-separated from $\mathbf{Y}$ in $G'_{\overline{x}}$, the graph obtained from $G'$ by cutting all arcs incoming to $X$. This implies $P(\mathbf{y}|w, do(x)) = P(\mathbf{y}|do(x))$. If it does, $P(\mathbf{y}|w, do(x))$ is identifiable iff $P(\mathbf{y}, w|do(x))$ is. In the former case, we are done, since if $P(\mathbf{y}|do(x))$ is not identifiable, then $P(\mathbf{Y}_x|x')$ isn't either. This is because results in [10] imply $P(\mathbf{Y}_x = \mathbf{y}|x')$ is identifiable iff either $P(\mathbf{y}|do(x))$ or $P(\mathbf{Y}_x = \mathbf{y}, X = x')$ is.

In the latter case there is a hedge $H$ [9] for $P(\mathbf{y}, w|do(x))$ in $G'$. In this case, it is possible to convert two counterexample models witnessing non-identification of $P(\mathbf{y}, w|do(x))$ into two counterexample models witnessing non-identification of $P(\mathbf{Y}_x = \mathbf{y}|x')$ by taking the values of $X$ in the new models to be the Cartesian product of the values of $X$ and $W$ in the old models (modifying the functions determining $X$ appropriately).

Next, we show that the estimand expression is correct.

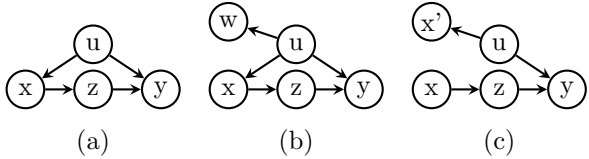

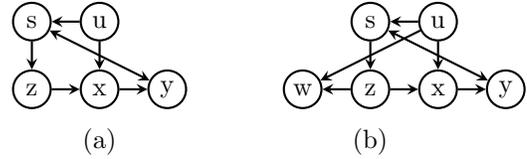

Figure 1: (a) A causal diagram $G$. (b) A graph $G'$ (from Theorem 1) such that $P(\mathbf{v}), G \vdash_{id} P(Y_x = y|x')$ iff $P(\mathbf{v}), G' \vdash_{id} P(y|w, do(x))$. (c) The counterfactual graph for $P(Y_x = y|x')$ in $G$.

Figure 3: (a) A causal diagram $G$ where $P(y|do(x))$ is identifiable (as $\frac{\sum_s P(y,x|z,s)P(s)}{\sum_s P(x|z,s)P(s)}$), but $P(Y_x|x')$ is not ($U$ is unobserved). (b) A graph $G'$ derived from $G$ using Theorem 1 where a new node $W$ has the same parents as $X$ but no children. This graph illustrates the difficulties of identifying $P(y|w, do(x))$.

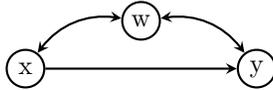

Figure 2: A causal diagram where $P(y|do(x))$ is identifiable, but $P(y|w, do(x))$ is not.

First, we note that since the counterfactual graph is just a causal diagram (with some nodes possibly being copies of each other), the estimand of $P(\mathbf{y}|w, do(x))$ is correct for $P(\mathbf{Y}_x = \mathbf{y}|x')$ up to variable renaming of certain copies. But $W$ is just a variable copy of $X$, thus we are justified in replacing values $w$ by $x'$.　□

We illustrate the application of Theorem 1 by considering the graph $G$ in Fig. 1 (a). The query $P(Y_x = y|x')$ is identifiable by considering $P(y|w, do(x))$ in the graph $G'$ shown in Fig. 1 (b), while the counterfactual graph for $P(Y_x = y|x')$ is shown in Fig. 1 (c). Identifying $P(y|w, do(x))$ in $G'$ using the algorithm in [8], we get $\sum_z P(z|x) \sum_x P(y|z, w, x) P(w, x)/P(w)$. Replacing $w$ by $x'$ yields the expression $\sum_z P(z|x) \sum_x P(y|z, x', x) P(x', x)/P(x')$. Note that $P(y|z, x', x)$ equals 0 for any value of $x$ other than $x'$. Simplifying we get that our query $P(Y_x = y|x')$ is equal to $\sum_z P(y|z, x') P(z|x)$.

Theorem 1 states that identifying $P(\mathbf{Y}_x = \mathbf{y}|x')$ is equivalent to identifying conditional causal effects like $P(\mathbf{y}|w, do(x))$. Conditional causal effects can be more difficult to identify compared to ordinary causal effects $P(\mathbf{y}|do(x))$. For instance, in Fig. 2, $P(y|do(x))$ is identifiable, while $P(y|w, do(x))$ is not, since conditioning on $W$ opens a confounding path between $X$ and $Y$. As a more complex example, consider the graph shown in Fig. 3 (a), where we are interested in identifying $P(Y_x = y|x')$. Note that in this graph, $P(y|do(x))$ is identifiable, while $P(y|w, do(x))$ in the graph shown in Fig. 3 (b), arising from the application of Theorem 1, is not. The reason is that conditioning on $W$ opens a confounding path $X \leftrightarrow S \leftrightarrow Y$.

### 3.1 Corollaries

We will discuss two important consequences of Theorem 1. The first is that the expression for $P(\mathbf{Y}_x = \mathbf{y}|x')$ can be derived from the expression for $P(\mathbf{y}|do(x))$ if the Backdoor Criterion [4] holds for some set $\mathbf{Z}$ (this criterion entails "conditional ignorability" in the language of [7]).

**Corollary 1 (Backdoor Criterion for ETT)** *If a set $\mathbf{Z}$ satisfies the Backdoor Criterion relative to $(X, \mathbf{Y})$, then $P(\mathbf{Y}_x = \mathbf{y}|x')$ is identifiable and equal to $\sum_\mathbf{z} P(\mathbf{y}|\mathbf{z}, x) P(\mathbf{z}|x')$.*

*Proof:* If the backdoor criterion holds in $G$, it also holds in $G'$ where $W$ is added. By construction, $Y \perp\!\!\!\perp W | \mathbf{Z}$ in $G'_{\overline{\mathbf{x}}}$. Thus $P(\mathbf{y}, w|do(x)) = \sum_\mathbf{z} P(\mathbf{y}|\mathbf{z}, do(x)) P(w, \mathbf{z}) = \sum_\mathbf{z} P(\mathbf{y}|\mathbf{z}, x) P(w, \mathbf{z})$. Thus, $P(\mathbf{y}|w, do(x)) = P(\mathbf{y}, w|do(x))/P(w) = \sum_\mathbf{z} P(\mathbf{y}|\mathbf{z}, x) P(\mathbf{z}|w)$. The conclusion follows.　□

This corollary is important in practice because adjustment for covariates is valid only if $\mathbf{Z}$ satisfies the Backdoor Criterion.

The second corollary concerns the so called additive intervention operation, denoted by $do(x' + q)$, which represents adding a quantity $q$ to a treatment variable $X$ that is currently at level $X = x'$. Clearly, the effect of the increment on $Y$ is given by the ETT expression $P(Y_x|x')$, with $x = x' + q$ which we may denote as $ETT(Y; x, x' + q)$. Likewise, the policy $do(X + q)$ of adding $q$ to $X$ *regardless* of its current value calls for evaluating the counterfactual expression $P(Y_{X+q} = y)$, where X is a variable. This latter quantity can be estimated as follows

$$P(y|do(X + q)) = \sum_{x'} P(Y_{X+q} = y|x') P(x') =$$

$$\sum_{x'} P(Y_{x'+q} = y|x')P(x') = E_{x'}ETT(Y; x', x' + q) \quad (1)$$

**Corollary 2** *The effect of the additive intervention $do(X+q)$ is identified if ETT is identified and, moreover, its estimand is given by Equation 1.*

Clearly, this result generalizes to any modifier $do(g(x))$ where $g(x)$ is an arbitrary function of $x$. Moreover, the effect of the conditional policy $do(g(x)$ if $X = x)$ can be assessed using the estimand given in Theorem 1.

It is interesting to note that, in general, the analysis of additive operators cannot be reduced to that of fixing operators, i.e., causal effects. In Figure 3, for example, the effect of fixing $X$ at $x' + q$ is identified while that of increasing $X$ from $x'$ to $x' + q$ is not. The reason is that the operator $do(x' + q)$ does not make use of the information provided by the observation $X = x'$, which may be significant. For example, a patient with a low serum level may react differently to increasing that level than a patient whose serum levels are already high.

Finally, we can use Theorem 1 to establish a simple graphical criterion which characterizes identification of ETT queries in terms of the original causal diagram, rather than the counterfactual graph. Before doing so, we introduce some helpful terminology. A C-component [12] of a latent projection is a maximal set of nodes pairwise connected by bidirected paths. Each causal diagram has a unique partition into C-components. A Q-factor [13] for a C-component in $G$ consisting of a set $\mathbf{Y}$, denoted by $Q[\mathbf{Y}]_{\mathbf{v}}^G$, is defined as $P(\mathbf{y}|do(\mathbf{v} \setminus \mathbf{y}))$, where $\mathbf{V}$ is the set of all observable nodes. Note that in this notation, $\mathbf{v}$ denotes the set of values to which variables are fixed, (or whose probabilities we are trying to calculate if the variables are in $\mathbf{Y}$). If the set $\mathbf{v}$ in the subscript of a Q-factor changes, the corresponding probability will also change. We will drop $G$ if it is obvious. It is known [13] that $Q[\mathbf{Y}]_{\mathbf{v}}^G$ is identifiable and equal to $\prod_i P(y^i|y_{\pi}^{(i-1)})$, where $Y^i$ are elements of $\mathbf{Y}$, $Y_{\pi}^{(i-1)}$ are elements preceding $Y^i$ in some fixed topological ordering $\pi$ consistent with the causal diagram, and values appearing in the above expression are consistent with the subscript $\mathbf{v}$. A C-forest [9] is a graph with nodes forming a C-component, such that every node has at most one observable child. The set of nodes in a C-forest without children is called its root set. We will denote graphs consisting of a vertex subset $\mathbf{X}$ of $G$ as $G_{\mathbf{x}}$. We are now ready to prove the main result of this section.

**Theorem 2** $P(\mathbf{v}), G \vdash_{id} P(\mathbf{Y}_x = \mathbf{y}|x')$ *if and only if there is no bidirected path from $X$ to a child of $X$ in $G_{an(\mathbf{y})}$. Moreover, if there is no such bidirected path, the estimand for $P(\mathbf{Y}_x = \mathbf{y}|x')$ is equal to $\sum_{\mathbf{v}\setminus(\mathbf{y}\cup\{x\})} \frac{P(an(\mathbf{y}))}{Q*P(x')} Q'$, where $Q$ is the Q-factor of $G_{an(\mathbf{y})}$ corresponding to the C-component containing $X$, and $Q'$ is that same Q-factor with all occurrences of $x$ are replaced by $x'$.*

*In other words, the estimand for $P(\mathbf{Y}_x = \mathbf{y}|x')$ is almost the same as the estimand for $P(\mathbf{y}|do(x))$ (from [13]), except rather than summing out $x$ from the expression for $Q'$, we replace $x$ by $x'$ in $Q'$, and divide by $P(x')$.*

*Proof:* See the extended technical report [11]. □

Theorem 2 says that in order to identify $P(\mathbf{Y}_x|x')$, it must be the case that the graph ancestral to $\mathbf{Y}$ contains no bidirected paths from $X$ to a child of $X$, and this effect is then identifiable by taking the product of the Q-factors of this graph, replacing all occurrences of the value $x$ in the Q-factor corresponding to the C-component containing X with the value $x'$, and marginalizing and conditioning on the appropriate variables to get the conditional distribution of interest.

Note the difference between this result and the one in [13]. Tian noted that the effect of a singleton $do(x)$ on a subset $\mathbf{Y}$ of $\mathbf{V}$ cannot be fully characterized by bidirected paths to a child of $X$ in $G_{an(\mathbf{y})}$ because such a path may go through a node ancestral to $X$ (and so this path isn't ancestral to $\mathbf{Y}$ if we cut incoming arrows to $X$). This abnormality cannot occur in our graph, since every node ancestral to $X$ in the original graph will be ancestral to the observable copy of $X$ in the counterfactual graph. In particular, in the graph shown in Fig. 3, $P(y|do(x))$ is identifiable but $P(Y_x|x')$ is not, since as we saw earlier, conditioning on $W$, creates confounding paths between $X$ and $Y$. Thus, in some sense the effect of treatment on the treated is an easier quantity to characterize (but not identify) than singleton effects.

We illustrate the application of Theorem 2 with an example. Consider the graph shown in Fig. 1 (a), where $X, Y, Z$ are observable, $U$ is latent, and we are interested in $P(Y_x = y|x')$. Since $U$ is unobserved, we can drop its node from the graph and consider the path from $X$ to $Y$ through $U$ as a single bidirected edge. Since $X$ has no bidirected paths to its descendant $Z$ in the latent projection, this query is identifiable. There are two C-components in Fig. 1 (a), $\{X, Y\}$ and $\{Z\}$. The Q-factor for the first C-component is $P(x, y|do(z)) = P(y|z, x)P(x)$, while the Q-factor for the second C-component is $P(z|do(y, x)) = P(z|do(x)) = P(z|x)$. Applying Theorem 2, we obtain the following identity:

$P(Y_x = y|x') = \left(\sum_z P(y|z,x')P(x')P(z|x)\right)/P(x') = \sum_z P(y|z,x')P(z|x)$. Note that this expression agrees with the expression we obtained using Theorem 1.

## 4 Experimental Evaluation of ETT and the Effect of Multiple Treatments

While ETT cannot in general be estimated from experimental studies (see counterexample in Figure 4 (a)), background knowledge can facilitate such estimation. For example, if the treatment is known to be binary, we immediately have: $P(Y_x = y) = P(Y_x = y|x')P(x') + P(y|x)P(x)$ which permits us to express ETT in term of the causal effect $P(Y_x = y)$. For non-binary treatments, the causal diagram may provide the knowledge needed.

To this end, we will generalize the queries we consider to be of the form $P(Y_\mathbf{x}^1 = y^1, ..., Y_\mathbf{x}^k = y^k|\mathbf{x}')$, where $\{Y^1, ..., Y^k\} = \mathbf{Y}$, and the set $\mathbf{X}$ is rule 3 minimal. In other words, there is no element $X \in \mathbf{X}$ such that rule 3 of do-calculus [4] applies to $X$, resulting in $P(\mathbf{y}|do(\mathbf{x})) = P(\mathbf{y}|do(\mathbf{x} \setminus \{x\}))$. This condition is simply a generalization of the requirement in the previous section that $X$ be ancestral to $\mathbf{Y}$. Since we consider effects of multiple "treatment" variables, we want to require that they are all causally relevant to $\mathbf{Y}$. We will abbreviate these queries as $P(\mathbf{Y}_\mathbf{x} = \mathbf{y}|\mathbf{x}')$, and call them the effects of multiple treatments on the treated.

The multiple treatment case is more complex in that rather than handling identification from $P(\mathbf{v})$ directly, it is more convenient to characterize identification from all possible interventional distributions first, that is from the set $P_* = \{P_\mathbf{x}(\mathbf{v} \setminus \mathbf{x}) | \mathbf{X} \subseteq \mathbf{V}\}$.

We first give a theorem which characterizes a set of causal diagrams where ETT is definitely not identifiable in general.

**Theorem 3** *Let $P(\mathbf{Y}_\mathbf{x} = \mathbf{y}|\mathbf{x}')$ be as above, and $\mathbf{X}$ be partitioned into sets $\mathbf{Z}$, $\mathbf{W}$, such that every element $X$ of $\mathbf{Z}$ has a backdoor path to $\mathbf{Y}$ in $G_{\overline{\mathbf{x} \setminus \{x\}}}$, and every element $X$ of $\mathbf{W}$ does not. Then $P_*, G \not\vdash_{id} P(\mathbf{Y}_\mathbf{x} = \mathbf{y}|\mathbf{x}')$ if there exists a C-component $F$ in $G_{\underline{\mathbf{z}}\overline{\mathbf{w}}}$ ancestral to $\mathbf{Y} \cup \mathbf{Z}$ such that either*

- *There is $W \in \mathbf{X}$ which has directed paths to both $\mathbf{Y}$ and $\mathbf{Z}$ in $G_{\underline{\mathbf{z}}\overline{\mathbf{w}}}$, such that the first node (after $W$) in both paths is in $F$.*

- *There is $Z \in \mathbf{Z}$ which is both a member of $F$, and would have had a directed path to $\mathbf{Y}$ in $G_{\underline{\mathbf{z}}\overline{\mathbf{w}}}$ with first node (after $Z$) on the path in $F$, but for the removal of it's outgoing arcs.*

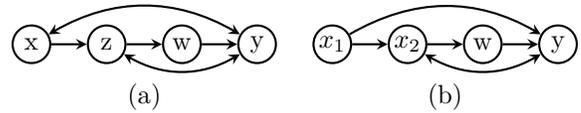

Figure 4: (a) $P(Y_x|x')$ is not identifiable from $P_*$. (b) $P(Y_{x_1,x_2}|x_1', x_2')$ is not identifiable from $P_*$.

*Proof:* See the extended technical report [11]. □

We illustrate how Theorem 3 can be used to conclude non-identification with two examples. Consider graphs in Fig. 4 (a) and (b). We already know from Theorem 2 that $P(Y_x|x')$ is not identifiable from $P(\mathbf{v})$ in Fig. 4 (a) since there exists a bidirected path from $X$ to $Z$. In fact, even if we are allowed to use all interventional distributions in $P_*$, $P(Y_x|x')$ remains non-identifiable in this graph due to Theorem 3. Since $X$ has a backdoor path to $Y$, it belongs to the set $\mathbf{Z}$ in the statement of Theorem 3. But there exists a C-component ($\{X,Y,Z\}$) ancestral to $\{X,Y\}$ in $G_{\underline{x}}$ such that $X$ both belongs to it, and has a path to $Y$ through the C-component but for the removal of the outgoing arc in $G_{\underline{x}}$. Similarly, $P(Y_{x_1,x_2} = y|x_1', x_2')$ is not identifiable in Fig. 4 (b). In this graph $\mathbf{W} = \{X_1\}$, and $\mathbf{Z} = \{X_2\}$, and there exists a C-component in $G_{\overline{x_1}\underline{x_2}}$, specifically $\{X_2, Y\}$, which is ancestral to $\{Y, X_2\}$, and there is an element of $\mathbf{W}$, namely $X_1$, which is a parent of this C-component and has a frontdoor path to both $Y$ and $X_2$ through this C-component.

As stated, Theorem 3 is not a complete result. We are going to argue that whenever it fails to hold, the corresponding treatment on the treated query is identifiable. The first step to showing this is the construction of a counterfactual graph. Fortunately, for queries where Theorem 3 fails, the counterfactual graph is relatively easy to construct from $G_{\underline{\mathbf{z}}\overline{\mathbf{w}}}$.

**Theorem 4** *Let $\gamma = P(\mathbf{Y}_\mathbf{x} = \mathbf{y}|\mathbf{x}')$ and $G$ be such that Theorem 3 does not hold, and $\mathbf{X}$ be partitioned into sets $\mathbf{Z}$, $\mathbf{W}$, such that every element $X$ of $\mathbf{Z}$ has a backdoor path to $\mathbf{Y}$ in $G_{\overline{\mathbf{x} \setminus \{x\}}}$, and every element $X$ of $\mathbf{W}$ does not. Then the counterfactual graph $G_\gamma$ can be obtained from $G_{\underline{\mathbf{z}}\overline{\mathbf{w}}}$ by the following operations:*

- *Split any node $W \in \mathbf{W}$ that contains frontdoor paths to $\mathbf{Y}$ and $\mathbf{Z}$ in $G_{\overline{\mathbf{x}}}$ into $W_1$ and $W_2$, where the former inherits frontdoor paths to $\mathbf{Y}$, and the latter inherits frontdoor paths to $\mathbf{Z}$.*

- *Copy any node $Z \in \mathbf{Z}$ that contains frontdoor paths to $\mathbf{Y}$ in $G_{\overline{\mathbf{x}}}$ such that the new copy $Z'$ contains all outgoing arrows of $Z$ in $G$.*

- *Remove all non-ancestors of $\mathbf{Y}$ and $\mathbf{Z}$.*

*Proof:* See the extended technical report [11]. □

We are finally ready to show that if Theorem 3 does not hold for $G$ and $\gamma$, then $\gamma$ is identifiable.

**Theorem 5** *Let $\gamma = P(Y_x = y|x')$ and $G$ be such that Theorem 3 does not hold, and $X$ be partitioned into sets $Z$, $W$, such that every element $X \in Z$ has a backdoor path to $Y$ in $G_{\overline{x \setminus \{x\}}}$, and every element $X$ of $W$ does not. Then $P_*, G \vdash_{id} \gamma$ and $\gamma$ is equal to*

$$\frac{\sum_{v' \setminus (y \cup z)} \prod_i Q[S_i]^{G_\gamma}_{v_\gamma}}{\sum_{v' \setminus z} \prod_i Q[S_i]^{G_\gamma}_{v_\gamma}}$$

*where $S_i$ is the set of C-component of $G_\gamma$, $V'$ are the observable unfixed nodes in $G_\gamma$, and $v_\gamma$ is defined such that free variables in the product are consistent with value assignments in $\gamma$.*

*Proof:* See the extended technical report [11]. □

Identification of $\gamma$ in terms of $P(\mathbf{v})$ can be ensured by using existing identification algorithms [12] to check that each Q-factor in the expression in Theorem 5 is identifiable from $P(\mathbf{v})$. In fact, since Theorem 5 ensures there are no conflicts among variable assignments in the Q-factors, it is possible to generalize Theorem 1 and Theorem 2 to this case.

**Theorem 6** *Let $\gamma = P(Y_x = y|x')$ and $G$ be such that Theorem 3 does not hold, and $X$ be partitioned into sets $Z$, $W$, such that every element $X$ of $Z$ has a backdoor path to $Y$ in $G_{\overline{x \setminus \{x\}}}$, and every element $X$ of $W$ does not. Then $P(\mathbf{v}), G \vdash_{id} \gamma$ if and only if $P(\mathbf{v}), G \vdash_{id} P(y|z, do(w))$.*

*Moreover, we can obtain the expression for $\gamma$ by running the **IDC** algorithm [8] on $P(y|z, do(w))$ and $P(\mathbf{v})$ in $G$, and replacing the values $x$ by values $x'$ in expressions for Q-factors $Q[S]^G$, for all nodes in $S$ which are also in $Z$, or for those nodes in $X$ which are parents of $S$ and have a frontdoor path to $Z$ through $S$ in $G_{\overline{z}\overline{w}}$.*

*Proof:* See the extended technical report [11]. □

Note that Theorem 6 is consistent with the expression obtained in the previous section, if we express $P(y|x)$ as $\sum_w P(y|w, x)P(w|x)$ first. In general, in order to render Theorem 6 consistent with previous results, **IDC** requires a slight modification where degenerate inputs of the form $P(\mathbf{y}|\mathbf{z})$ which contain no interventions are not returned as is, but instead are returned in the form

$$\frac{\sum_{\mathbf{v} \setminus (\mathbf{y} \cup \mathbf{z})} \prod_i P(s_i|do(\mathbf{v} \setminus s_i))}{\sum_{\mathbf{v} \setminus (\mathbf{z})} \prod_i P(s_i|do(\mathbf{v} \setminus s_i))}$$

where $S_i$ are the C-components of the corresponding graph, and each term $P(s_i|do(\mathbf{v} \setminus s_i))$ is identifiable [13] as $\prod_{x_j \in s_i} P(x_j|x_\pi^{(j-1)})$, where $X_\pi^{(j-1)}$ is the set of nodes preceding $X_j$ in some topological order $\pi$ consistent with the graph.

We illustrate Theorems 4, 5 and 6 with an example. Consider the graph shown in Fig. 5 (a). It's easy to see that Theorem 3 does not apply to the query $P(Y_{x_1,x_2} = y|x'_1, x'_2)$ in this graph. Furthermore, $X_2$ has a backdoor path to $Y$ and $X_1$ does not. Thus, the graph from which the counterfactual graph is to be constructed is $G_{\overline{x_1}x_2}$. We add a copy of $X_2$ with outgoing arrows only (e.g. an arrow to $W$). Furthermore, since $X_1$ contains frontdoor paths to both $X_2$ and $Y$ in $G_{\overline{x_1},x_2}$, we copy $X_1$ as well, and have the copy corresponding to the event $do(x_1)$ inherit all outgoing arrows corresponding to frontdoor paths to $Y$ (in this case there's only one such arrow – to $W$), and the copy corresponding to the event $X_1 = x'_1$ inherit all outgoing arrows corresponding to frontdoor paths to $X_2$ (in this case there's only one such arrow – to $X'_2$). The resulting counterfactual graph is shown in Fig. 5 (b). Using Theorem 5, we can now identify $P(Y_{x_1,x_2} = y|x'_1, x'_2)$ using the Q-factors corresponding to the counterfactual graph as equal to

$$\frac{\sum_w P_{x'_1,w}(y, x'_2) P_{x_1,x_2}(w)}{\sum_{w,y} P_{x'_1,w}(y, x'_2) P_{x_1,x_2}(w)} \quad (2)$$

Furthermore, each of the Q-factors is identifiable in the original causal diagram. Specifically, $P_{x_1,x_2}(w) = P(w|x_1, x_2)$, and $P_{x'_1,w}(y, x'_2) = P(y|w, x'_2, x'_1)P(x'_2|x'_1)$. Thus, $P(Y_{x_1,x_2} = y|x'_1, x'_2) =$

$$\frac{\sum_w P(y|w, x'_2, x'_1) P(x'_2|x'_1) P(w|x_1, x_2)}{\sum_{w,y} P(y|w, x'_2, x'_1) P(x'_2|x'_1) P(w|x_1, x_2)}$$

which is just equal to $\sum_w P(y|w, x'_2, x'_1) P(w|x_1, x_2)$ after canceling. Note that this expression coincides with the output of **IDC** identifying $P_{x_1}(y|x_2)$ in $G$, where the values $x_1, x_2$ in expressions for the Q-factors containing $X_2$ are replaced with values $x'_1, x'_2$, as we would expect from Theorem 6. We note also that the addition of a bidirected arc from $X_1$ to $X_2$ in Fig. 5 (a) renders the query $P(Y_{x_1,x_2} = y|x'_1, x'_2)$ identifiable from $P_*$ (by 2) but not from $P(\mathbf{v})$, which means Theorem 5 does not entail Theorem 6 in all graphs.

## 5 Conclusions

We have formulated the problem of estimating the counterfactual probability $P(Y_x = y|x')$ as a problem of estimating a distribution under conditional intervention. We have shown that if only a single in-tervention is performed, identification of such queries in terms of the observable distribution has a simple characterization in terms of the causal diagram. This characterization permits the evaluation of additive interventions (e.g., $do(X + q)$), and produces immediate estimands for ETT whenever causal effects can be estimated by covariate adjustment.

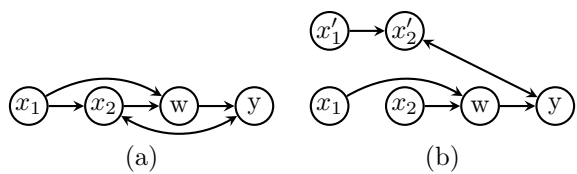

Figure 5: (a) A causal diagram where $P(Y_{x_1,x_2} = y|x'_1, x'_2)$ is identifiable from $P_*$ (and $P(\mathbf{v})$). (b) A counterfactual graph for $P(Y_{x_1,x_2} = y|x'_1, x'_2)$ created using Theorem 4.

We further showed that, for multiple interventions it is possible to characterize identification in terms of experimental distributions by using the original causal diagram, rather than having to construct additional graphical representations of multiple hypothetical worlds implicit in such queries [10]. Finally, we show that identified expressions for the queries we consider can be obtained by value replacement operations performed on identified expressions of ordinary conditional effects of the form $P(\mathbf{y}|\mathbf{w}, do(\mathbf{x}))$ which contain no value conflicts.

We speculate that, since ETT measures whether an actor should regret having taken an action do(x) (prior to observing the outcome), its identification should play a role in robot learning applications. *Regret* is the fundamental mechanism behind the evolution of ethical behavior in humans; it should therefore govern the development of collaborative societies of robots.

**Acknowledgments**

The authors thank anonymous reviewers for helpful comments. This work was supported in part by the Office of Naval Research (contract/grant number: 1053729), and NIH grant #R37AI032475.

**References**

[1] J. J. Heckman. *Randomization and social policy evaluation.* In: Evaluations: Welfare and Training Programs, Cambridge, MA: Harvard University Press, 1992.

[2] J. Neyman. Sur les applications de la thar des probabilities aux experiences Agaricales: Essay des principle. Excerpts reprinted (1990) in English. *Statistical Science*, 5:463–472, 1923.

[3] Judea Pearl. *Probabilistic Reasoning in Intelligent Systems.* Morgan and Kaufmann, San Mateo, 1988.

[4] Judea Pearl. *Causality: Models, Reasoning, and Inference.* Cambridge University Press, 2000.

[5] Thomas Richardson and Peter Spirtes. Ancestral graph markov models. *Annals of Statistics*, 30:962–1030, 2002.

[6] James M. Robins, VanderWeele Tyler J., and Thomas S. Richardson. Comment on causal effects in the presence of non compliance: a latent variable interpretation by antonio forcina. *METRON*, LXIV(3):288–298, 2006.

[7] D. B. Rubin. Estimating causal effects of treatments in randomized and non-randomized studies. *Journal of Educational Psychology*, 66:688–701, 1974.

[8] Ilya Shpitser and Judea Pearl. Identification of conditional interventional distributions. In *Uncertainty in Artificial Intelligence*, volume 22, 2006.

[9] Ilya Shpitser and Judea Pearl. Identification of joint interventional distributions in recursive semi-markovian causal models. In *21st National Conference on Artificial Intelligence*, 2006.

[10] Ilya Shpitser and Judea Pearl. What counterfactuals can be tested. In *23rd Conference on Uncertainty in Artificial Intelligence*. Morgan Kaufmann, 2007.

[11] Ilya Shpitser and Judea Pearl. Effects of treatment on the treated: Identification and generalization. Technical Report R-349-L, Cognitive Systems Laboratory, University of California, Los Angeles, 2009.

[12] Jin Tian. *Studies in Causal Reasoning and Learning.* PhD thesis, Department of Computer Science, University of California, Los Angeles, 2002.

[13] Jin Tian and Judea Pearl. A general identification condition for causal effects. In *18th National Conference on Artificial Intelligence*, pages 567–573, 2002.

[14] T. S. Verma and Judea Pearl. Equivalence and synthesis of causal models. Technical Report R-150, Department of Computer Science, University of California, Los Angeles, 1990.